\documentclass[twocolumn,prl,amsmath,amssymb,floatfix]{revtex4} 
\usepackage{graphicx} 
\usepackage{epsfig} 
\usepackage{ae}
\usepackage{aecompl}

\begin{document}

\title{ Field Theoretic Study of Bilayer Membrane Fusion:
II. Mechanism of a Stalk-Hole Complex}

\author{K.\ Katsov}
\affiliation{Materials Research Laboratory, University of California, 
Santa Barbara CA 93106}
\author{M. M{\"u}ller}
\affiliation{Institut f{\"u}r Theoretische Physik, Friedrich-Hund-Platz 1, 
37077 G{\"o}ttingen, Germany}
\author{M. Schick}
\affiliation{Department of Physics, University of Washington,  Box
  351560, Seattle, WA 98195-1560}

\date{July 19, 2005} 
\begin{abstract} We use self-consistent field theory 
to determine structural and energetic properties of intermediates and 
transition states involved in bilayer membrane fusion. In particular, we 
extend our original calculations from those of the standard hemifusion 
mechanism, which was studied in detail in the first paper of this series 
\cite{Katsov04}, to consider a possible alternative to it. This mechanism 
involves non-axial stalk expansion, in contrast to the axially symmetric 
evolution postulated in the classical mechanism. Elongation of the initial 
stalk facilitates the nucleation of holes and leads to destabilization of the 
fusing membranes via the formation 
of a stalk-hole complex. We study properties of this complex in detail, 
and show how transient leakage during fusion, previously predicted and 
recently observed in experiment, should vary with system architecture and 
tension. We also show that the barrier to fusion in the alternative 
mechanism is lower than that of the standard mechanism by a few 
$k_BT$ over most of the relevant region of system parameters, so that this 
alternative mechanism is a viable alternative to the standard pathway.

\end{abstract}
\maketitle

\section{Introduction} The fusion of biological membranes is of great 
importance as it plays a central role {\em inter alia} in intracellular 
trafficking, exocytosis, and viral infection 
\cite{Blumenthal03,Jahn02,Lentz00,Mayer02,Skehel00,Tamm03}. Given this 
importance, it might be thought that its mechanism would be well understood, 
but in fact it is not. Perhaps the reason for this is that there is
an apparent dilemma at the heart of the fusion process. 
The vesicles, or bilayers, which are to be fused must be sufficiently stable
with respect to irreversible rupture in order to carry out  
their functions on a reasonably long time scale. It follows that it must be 
quite energetically expensive to create a long-lived, super-critical, hole in
such a membrane.
In other words, the free energy barrier to do so must be very large 
compared to the thermal fluctuation energy $k_BT$. As a consequence, almost 
all holes created by thermal fluctuations do not have sufficient energy to 
traverse this barrier, hence they simply shrink and reseal. However, it 
is inevitable that for fusion to occur, a long-lived hole must be created 
at some stage of the fusion pathway. The dilemma is that a bilayer can both 
be stable with respect to rupture and yet readily undergo fusion.

Some of the solution of this puzzle is in place. It is believed that 
fusion proteins locally expend energy to dehydrate both bilayers
in order to bring them in close proximity. This increases the free energy per 
unit area of the system, {\em i.e} puts the system under local stress. As a 
consequence, it is free energetically favorable for the system 
to undergo a transformation that 
results in a decrease of bilayer area.
In principle, this can be accomplished both by fusion and/or rupture, but
the proteins apparently catalyze  the fusion process exclusively.

The standard hemifusion mechanism, proposed twenty years ago by Kozlov and 
Markin \cite{Kozlov83}, assumes that thermal fluctuations permit the tails 
of lipids of the {\em cis} leaves, those of the apposing membranes which are 
closest to one another, to flip over and form an 
axially symmetric defect in the dehydrated region, denoted a stalk 
\cite{Chernomordik03}. Due to 
the tension, the newly joined cis layers recede so that the stalk expands 
radially preserving the axial symmetry, and transforms into a hemifusion
diaphragm -  
a {\em single} bilayer consisting of the two remaining {\em trans} leaves.
Only this single bilayer needs to be punctured by a hole in order 
that a fusion pore be formed and the fusion process completed.
This radial stalk expansion hypothesis, being in qualitative agreement with many
experimental observations, was essentially the only model of the
fusion process until recently.

In contrast to the hemifusion hypothesis, Monte Carlo simulations of bilayer
fusion \cite{Mueller03} showed that fusion can evolve through an 
alternative mechanism \cite{Noguchi01,Mueller02}, in which the stalk 
does not expand radially, but rather elongates in a worm-like fashion.
To distinguish the original axially symmetric stalk from the
elongated structure, we will call the former the ``classical stalk'' 
for the remainder of the paper.
Moreover, it was observed that the elongated stalk destabilizes the fusing
membranes by greatly enhancing the rate of hole formation in its vicinity.
Once such a hole is formed in one bilayer close to the elongated stalk, the
stalk encircles it completely forming a hemifusion diaphragm consisting of the
other, as yet intact, bilayer. Subsequent hole formation in this diaphragm
completes the fusion process.  
In a slightly different variant of this scenario, holes form in both bilayers  
near the stalk before the stalk has completely surrounded the first hole. 
Fusion is completed when the stalk surrounds both holes. This mechanism 
was also seen in recent Molecular Dynamics simulations 
\cite{Marrink03,Stevens03}. It was argued \cite{Mueller03} that the stalk 
lowers the free energy barrier to hole formation by decreasing the effective
line tension in that part of the hole in contact with the stalk. 

We shall denote the elongated stalk partially surrounding a hole as a 
{\em stalk-hole complex}. As we note below, this stalk-hole complex can decay,
i.e. evolve without further free energy cost, into a final fusion pore, so
that {\em this complex represents a potential transition state in the fusion
process}.  

A direct consequence of this alternative mechanism is that there can be 
transient leakage during fusion. Even though leakage is sometimes observed
during fusion experiments, it is usually attributed to the presence of fusion 
proteins which are known, for example, to initiate erythrocyte hemolysis 
\cite{Niles90}. However, the new mechanism predicts that transient 
leakage stems from the fusion pathway itself and should be
observable even during fusion of model membranes in the absence of fusion
proteins. Leakage during fusion in such systems has indeed been observed 
experimentally \cite{cevc99,evans02,lentz97}. 
In addition, it is predicted that this transient leakage should be {\em 
correlated} in space and time with fusion. Just such correlated leakage 
and fusion were recently observed experimentally by Frolov et al. 
\cite{Frolov03}. Fusion without detectable leakage is also observed, 
however \cite {Smit02,Spruce91,Tse93}.  We shall argue below that the 
seeming irregularity of leakage accompanying fusion can be explained
by the new mechanism. In particular, the extent of this transient leakage
depends both on the architecure of the amphiphiles as well as the tension
(stress) imposed on the membranes. By decreasing the spontaneous curvature of
the amphiphiles and/or reducing the membrane tension, the
leakage can be substantially reduced and even completely elliminated in some
cases. 

While simulations can reveal very clearly the process by which fusion 
occurs, it is very difficult to extract free energy barriers involved in the
process.  Moreover, it would be prohibitively expensive to study comprehensively
the fusion process in a range of parameters such as amphiphile
architecture and/or bilayer tension. 

In a previous paper \cite{Katsov04}, we employed self-consistent field theory 
(SCFT) to evaluate these barriers {\em assuming} that fusion took place 
via the standard, radially-expanding stalk and hemifusion mechanism. The 
system considered consisted of bilayers of $AB$ block copolymer, with 
fraction $f$ of the $A$ monomer, in a solvent of $A$ homopolymer. All 
polymers were characterized by the same polymerization index. Comparison 
of various properties of this system of block copolymer amphiphiles with 
those of membranes consisting of biological lipids permitted an estimate 
that free energies of a structure in the copolymer simulations were 2.5 times 
%MM: the factor 2.5 depends on the specific model -- for experimental systems
%it will be much smaller because of the higher density and chain length.
smaller than those of the corresponding structure in the biological 
system. We calculated the barrier to stalk formation in polymeric 
bilayers, and from it estimated that in membranes made of biological 
lipids, this barrier would not exceed 13$k_BT$. The larger barrier in the 
standard process is that associated with the radial expansion of the 
hemifusion diaphragm \cite{Kozlovsky002}, and we estimated this to be in 
the range of 25-63$k_BT$, depending upon the lipid architecture and 
membrane tension. Perhaps one of the most interesting results of this 
study was the following:  the range of variation in amphiphile 
architecture over which successful fusion can occur is {\em severely} 
restricted by the fact that the fusion process begins with the formation 
of a metastable, classical stalk. If $f$ is too large, 
corresponding to lipids with very small spontaneous curvature, stalks 
between bilayers are never metastable. On the other hand, if $f$ is too 
small, corresponding to lipids with larger negative spontaneous 
curvatures, {\em linear} elongated stalks became favorable which
destabilize the  
bilayers completely by causing a transition to an inverted hexagonal 
phase.  Thus in order for fusion to occur, the lipid composition of 
membranes must be tightly regulated. This conclusion also applies to 
fusion which proceeds via the new mechanism as it, too, begins with the 
formation of the classical stalk.

In this paper we apply SCFT methods to calculate the fusion barriers in 
this new, alternative, mechanism. We begin in Section~\ref{sec:hole} by calculating 
the free energy of an isolated hole in a single bilayer as a function of 
its radius, $R$, for bilayers under various tensions, $\gamma$ and 
consisting of diblocks of different architectural parameters, $f$.  The 
result is that, as expected, it is very expensive to create a hole in an 
isolated bilayer. In Section~\ref{sec:sh-complex} we turn to the calculation of the free 
energy of the stalk-hole complex. Because this complex is {\em not} 
axially symmetric, our task is much more difficult than our previous 
calculation of the barriers in the old hemifusion mechanism in which the 
intermediates were postulated to be axially symmetric. 
We accomplish our goal by 
constructing the non-axially symmetric intermediates from fragments of other 
excitations which do possess this symmetry, and therefore are more easily 
obtained. In particular, we show that structures related to an Inverted Micellar
Intermediate (IMI) play important role in this process.
We compare our results for the free energy barrier in the two
mechanisms and show that the barrier in the new 
one is indeed lower than that in the old, although the difference in most 
of the region of parameters in which fusion can occur successfully is not 
more than a few $k_BT$. Finally in Section~\ref{sec:disc} we discuss further the 
reason why the new mechanism is a favorable one. We trace it not only to 
the reduction of the line tension of a hole when nucleated next to a 
stalk, but also to the relatively low cost for the stalk to extend 
linearly. Consequently when a hole appears in the bilayer, a large 
fraction of its circumference can have its line tension reduced by the 
nearby presence of a stalk. We conclude with some comments on the
dependence of the rate of hole formation in a bilayer on the line
tension of the hole. We show that even modest changes in the effective line
tension of a hole due to the presence of the elongated stalk in the stalk-hole
complex can strongly affect the rate of hole formation, and hence the rate of
fusion. Such small changes in line tension, therefore, destabilize what 
were very stable bilayers and enable them to undergo fusion.

\section{Free Energy of a Hole in an Isolated Bilayer}
\label{sec:hole}
In this section we discuss the free energy of a circular hole in an isolated 
bilayer to show that the energy associated with formation of such a defect is
high, as is expected if isolated bilayers are stable. The SCFT calculation follows 
the lines described in our previous paper \cite{Katsov04}. It is 
straightforward within the SCFT to obtain the free energy, 
$\Omega_m(T,\Delta\mu,V, A )$, of a bilayer of area $ A$ at a temperature 
$T$ and a difference, $\Delta \mu=\mu_a-\mu_s$, of the bulk chemical 
potentials of the amphiphile and of the solvent. There is only one 
independent chemical potential as the system is assumed to be 
incompressible. The volume of the system is $V$. Similarly, we denote the 
free energy of the system without the bilayer, i.e., a homogeneous 
amphiphile solution, $\Omega_0(T,\Delta\mu,V)$.  The difference between 
these two free energies, in the thermodynamic limit of infinite volume, 
defines the excess free energy of the bilayer membrane:
\begin{equation}
\delta\Omega_m(T,\Delta\mu, A)\equiv
\lim_{V\rightarrow\infty}[\Omega_m(T,\Delta\mu,V, A)-
\Omega_0(T,\Delta\mu,V)].
\end{equation}
The excess free energy per unit area, in the thermodynamic limit of
infinite area, defines the lateral membrane tension
\begin{equation}
\gamma(T,\Delta\mu)\equiv\lim_{A\rightarrow\infty} 
[\delta \Omega_m(T,\Delta\mu,A)/ A].
\end{equation}
Changes in this tension $\gamma$ can be related to changes in the 
temperature and chemical potential by means of the Gibbs-Duhem equation
\begin{equation} 
{\rm d}\gamma(T,\Delta\mu)=-\delta s\ {\rm d}T-\delta
\sigma_a {\rm d}(\Delta\mu), 
\end{equation}
where $\delta s$ is the excess entropy per unit area, and $\delta\sigma_a$
is the excess number of amphiphilic molecules per unit area. This relation
shows that the chemical potential difference, $\Delta\mu$, can be used
to adjust the bilayer tension $\gamma$.

As discussed previously \cite{Katsov04}, it is also possible to introduce 
axially symmetric defects of a specified radius $R$ into the bilayer and 
to obtain the excess free energy of such structures.
The choice of model parameters was dictated by our original Monte
Carlo simulations \cite{Mueller03} and the details can be found in the first
paper of this series \cite{Katsov04}. 

%%%%%%%%%%%%%%%%%%%%%%%%%%%%%%%%%%%%%%%%
%%%%%%%%%%%%%%%%%%%%%%%%%%%%%%%%%%%%%%%%
%%%% FIGURE 1.
%%%%%%%%%%%%%%%%%%%%%%%%%%%%%%%%%%%%%%%%
%%%%%%%%%%%%%%%%%%%%%%%%%%%%%%%%%%%%%%%%
\begin{figure}[t]
\includegraphics[width=3in]{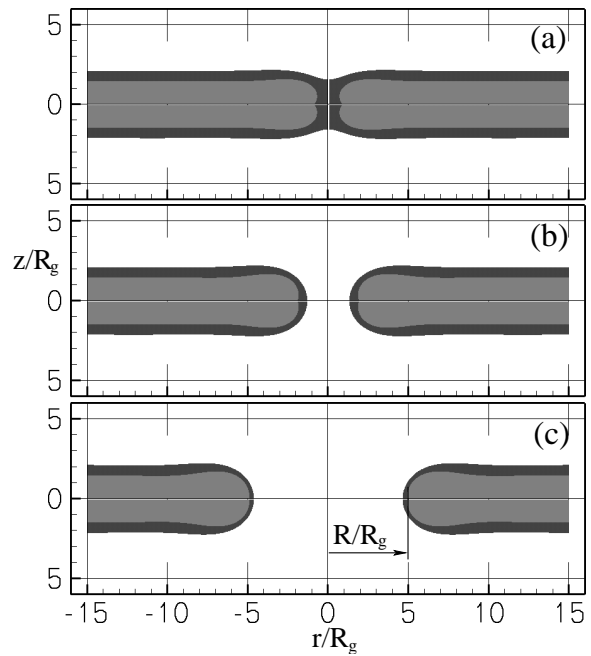} 
\caption{Density profiles of bilayers pierced by an isolated hole 
are shown for three different hole radii:
$R/R_g=$ 1, 2, and 5. Only the majority component is shown at each
point. Solvent segments are white. Hydrophilic and hydrophobic
segments of the amphiphile are shaded dark and light respectively.} 
\label{fig:profhole}%
\end{figure}
%%%%%%%%%%%%%%%%%%%%%%%%%%%%%%%%%%%%%%%%

Fig.~\ref{fig:profhole} shows the density distribution of hydrophobic (B) and
hydrophilic (A) segments in a bilayer with holes of different radii,
which are defined as the radial distance in the plane of symmetry 
to the $A/B$ interface, the point at which the volume fractions of A and
B monomers are equal. We find that qualitative features of this profile 
are not very
sensitive to the architectural parameter $f$ or tension $\gamma$: the rim of
the hole has a shape of a bulb which is typical whenever a flat bilayer
has an edge \cite{Matsen97,Netz97,Kasson04}. 

%%%%%%%%%%%%%%%%%%%%%%%%%%%%%%%%%%%%%%%%
%%%%%%%%%%%%%%%%%%%%%%%%%%%%%%%%%%%%%%%%
%%%%%%%%% FIG 2   
%%%%%%%%%%%%%%%%%%%%%%%%%%%%%%%%%%%%%%%%
%%%%%%%%%%%%%%%%%%%%%%%%%%%%%%%%%%%%%%%%
\begin{figure}[t]
\includegraphics[width=3in]{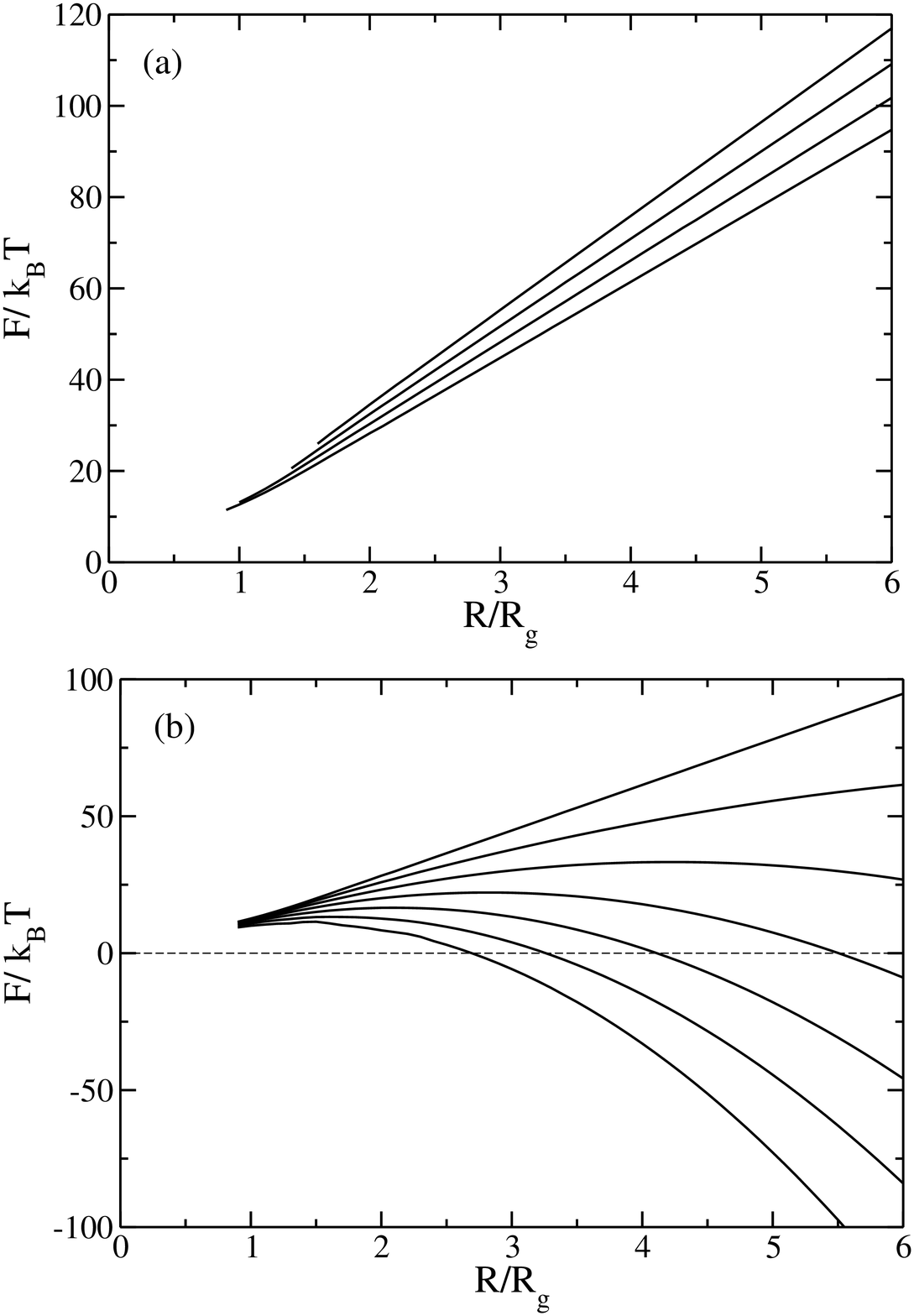}
\caption{(a) Free energy of a hole in an isolated bilayer 
as a function of $R/R_g$ at zero tension 
for various amphiphile architectures, $f$. From top to bottom the values
of $f$ are 0.29, 0.31, 0.33, and 0.35. (b) Same as above, but at fixed
$f=0.35$ and various tensions $\gamma/\gamma_0$. From top to bottom,
$\gamma/\gamma_0$ varies from 0.0 to 0.6 in increments of 0.1.}
\label{fig:holefree}
\end{figure}
%%%%%%%%%%%%%%%%%%%%%%%%%%%%%%%%%%%%%%%%

The free energy of such a hole in a bilayer is shown in
Fig.~\ref{fig:holefree}(a) as a function of its radius for a bilayer at zero  
tension and composed of amphiphiles of different architectural parameters, 
$f$.  One sees that under zero tension the free energy increases
essentially linearly with $R$ and the excess free energy of the hole can be
written as $2\pi \lambda_H(T,\Delta\mu,R)R$, where $\lambda_H$ is an effective
line tension. As one would expect, this line tension quickly asymptotes to a
constant value $\lambda_H(T,\Delta\mu)$ for sufficiently large $R$. For the
bilayer under zero tension composed of amphiphiles with $f=0.35$, we find
$\lambda_H R_g/kT=2.63$. To  
compare with analogous values for membranes, we convert this to the 
dimensionless ratio $\lambda_H/\gamma_0 d$, where $\gamma_0$ is the free 
energy per unit area of an interface between coexisting phases of bulk 
homopolymer $A$ and bulk homopolymer $B$, and $d$ is the thickness of the 
bilayer. From our previous work \cite{Mueller03}, we obtain 
$\gamma_0d^2/kT=65.3$, and $d/R_g=4.47$ so that $\lambda_H/\gamma_0 
d=0.18$. The analogous quantity can be calculated for membranes taking 
$\lambda_m=2.6\times 10^{-6}$~dynes \cite{Moroz97,Zhelev93}, 
$d_m=35.9\times10^{-8}$~cm \cite{Rand90}, and an oil-water tension of 
$\gamma_m=50$~dynes/cm, from which $\lambda_m/\gamma_m d_m=0.14$. 
Thus the line tensions we obtain are reasonable.

In Fig.~\ref{fig:holefree}(b) we show the effect of membrane tension on the
dependence of the hole free energy on radius $R$. One sees, as expected, that the 
free energy of the hole eventually decreases due to the elimination of 
stressed membrane area. For sufficiently large $R$, one expects the free energy
of the hole to be of the form
\begin{equation}
F_H(R)=2\pi \lambda_H R-\pi \gamma R^2,
\label{holefree}
\end{equation}
with $\gamma$ the imposed membrane tension. We verified that this form is
certainly adequate at  
large $R$. At smaller radii, which will be of interest to us, the 
coefficients $\lambda_H$ and $\gamma$ are, themselves, functions of $R$.  
One sees from this figure that under relevant tensions the maximum value of 
$F_H(R)$, which is required to form a critical size hole leading to
irreversible membrane rupture, is no less than about 16$k_BT$. In a bilayer 
composed of biological lipids, we can estimete that this would correspond to a
barrier of approximately 40$k_BT$. Thus, as stated, isolated bilayers are very
robust against rupture caused by  thermal excitation, and it is precisely this
stability that makes fusion difficult to understand.
 
We now turn to the calculation of the stalk-hole complex, which is a possible 
fusion intermediate. We will show that the barrier to fusion is much less 
than the barrier to create a hole in each of the two fusing bilayers in 
the absence of an elongated stalk.
 
\section{Free energy of the stalk-hole complex}
\label{sec:sh-complex}
\subsection{Immediately before formation of stalk-hole complex}
 
%%%%%%%%%%%%%%%%%%%%%%%%%%%%%%%%%%%%%%%%
%%%%%%%%%%%%%%%%%%%%%%%%%%%%%%%%%%%%%%%%
%%%%%%%%% FIG 3 
%%%%%%%%%%%%%%%%%%%%%%%%%%%%%%%%%%%%%%%%
%%%%%%%%%%%%%%%%%%%%%%%%%%%%%%%%%%%%%%%%

\begin{figure}[t]
\includegraphics[width=3in]{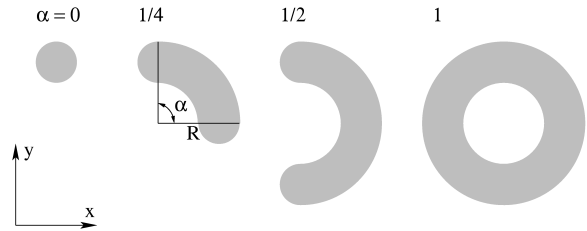}
\caption{Parametrization of the elongated stalk. Gray schematically shows
  location of the hydrophobic segments in the plane of symmetry between fusing
  bilayers. The arc radius $R$ corresponds to
  the radial distance to the outer hydrophilic/hydrophobic interface in the
  plane of symmetry.
  Values of the fractional arc angle $\alpha$, defined in the range $[0,1]$, are
  given at the top of each stalk configuration. Note that $\alpha=0$
  corresponds to the original stalk, whereas $\alpha=1$ corresponds
  to a family of structures reminiscent of the IMI (see also Fig.~\ref{fig:imi}).}
\label{fig:estalk}%
\end{figure}
%%%%%%%%%%%%%%%%%%%%%%%%%%%%%%%%%%%%%%%%

\begin{figure}[t]
\includegraphics[width=3in]{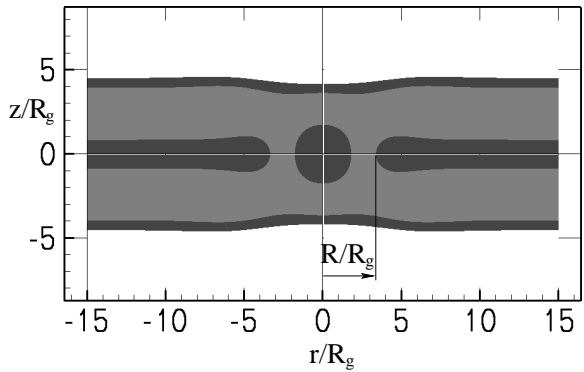} 
\caption{Density profile of an IMI. The amphiphiles are
characterized by $f=0.3$. The radius of the IMI, in units of the ideal radius of
gyration, $R/R_g$ is 3.4. Gray scale as in Fig.~\ref{fig:profhole}.}
\label{fig:imi}%
\end{figure}

Right after the formation of the initial (classical circular) stalk and
just before the formation of the stalk-hole complex, the stalk
elongates in a worm-like fashion. For the sake of simplicity, we assume that
in the $z=0$ (symmetry) plane this elongated structure has a shape of a
circular arc with a fractional angle, $0\leq\alpha\leq 1$, and radius 
$R$, as shown schematically in Fig.~\ref{fig:estalk}. With this choice of the
parameters, $\alpha=0$ corresponds to the classical stalk structure, whereas
for $\alpha=1$ there is a family of structures that are reminiscent of the
Inverted Micellar Intermediate (IMI), studied previously by Siegel
\cite{Siegel93}. Although the radii of the structures we  
consider are not necessarily small, we shall continue to refer to them as 
IMIs. A density profile of one such structure is shown in
Fig.~\ref{fig:imi}. Its radius $R$ is defined as the radial distance to the
furtherest point on the $z=0$ plane at which the densities  
of hydrophobic and hydrophilic segments are equal, and is shown in the 
figure. We denote its free energy $F_{IMI}(R)$. 
Note that the equilibrium IMIs considered by Siegel correspond to structures
with an optimal radius $R^*$, which minimizes $F_{IMI}(R)$.

%%%%%%%%%%%%%%%%%%%%%%%%%%%%%%%%%%%%%%%%%%%
%%%%%%%%%%%%%%%%%%%%%%%%%%%%%%%%%%%%%%%%%%%
%%%%%%%%%%%   Figure 4
%%%%%%%%%%%%%%%%%%%%%%%%%%%%%%%%%%%%%%%%%%%
%%%%%%%%%%%%%%%%%%%%%%%%%%%%%%%%%%%%%%%%%%%

%%%%%%%%%%%%%%%%%%%%%%%%%%%%%%%%%%%%%%%%%%%

In general, the elongated stalk will not form a complete IMI, 
that is $\alpha$ will be less than unity, 
so we approximate the free energy of the 
extended stalk in this configuration as
 \begin{equation}
 F_1(R,\alpha)=\alpha F_{IMI}(R)+F_S.
 \label{justbefore}
 \end{equation} 
The presence of the second term is due to the free energy of the end caps of 
the extended stalk (see Fig.~\ref{fig:estalk}). As these two ends together
form an axially symmetric  
stalk, the free energy of these ends is just the free energy of the classical 
stalk, $F_S$, which we have calculated previously \cite{Katsov04}. Note that
for the case $\alpha=1$, the above estimate is certainly an upper bound as the 
second term should be absent in that case.
 
%%%%%%%%%%%%%%%%%%%%%%%%%%%%%%%%%%%%%
%%%%%%%%%%%%%%%%%%%%%%%%%%%%%%%%%%%%
%%%%%      Figure 5
%%%%%%%%%%%%%%%%%%%%%%%%%%%%%%%%%%%
%%%%%%%%%%%%%%%%%%%%%%%%%%%%%%%%%%%

\begin{figure}[t]
\includegraphics[width=3in]{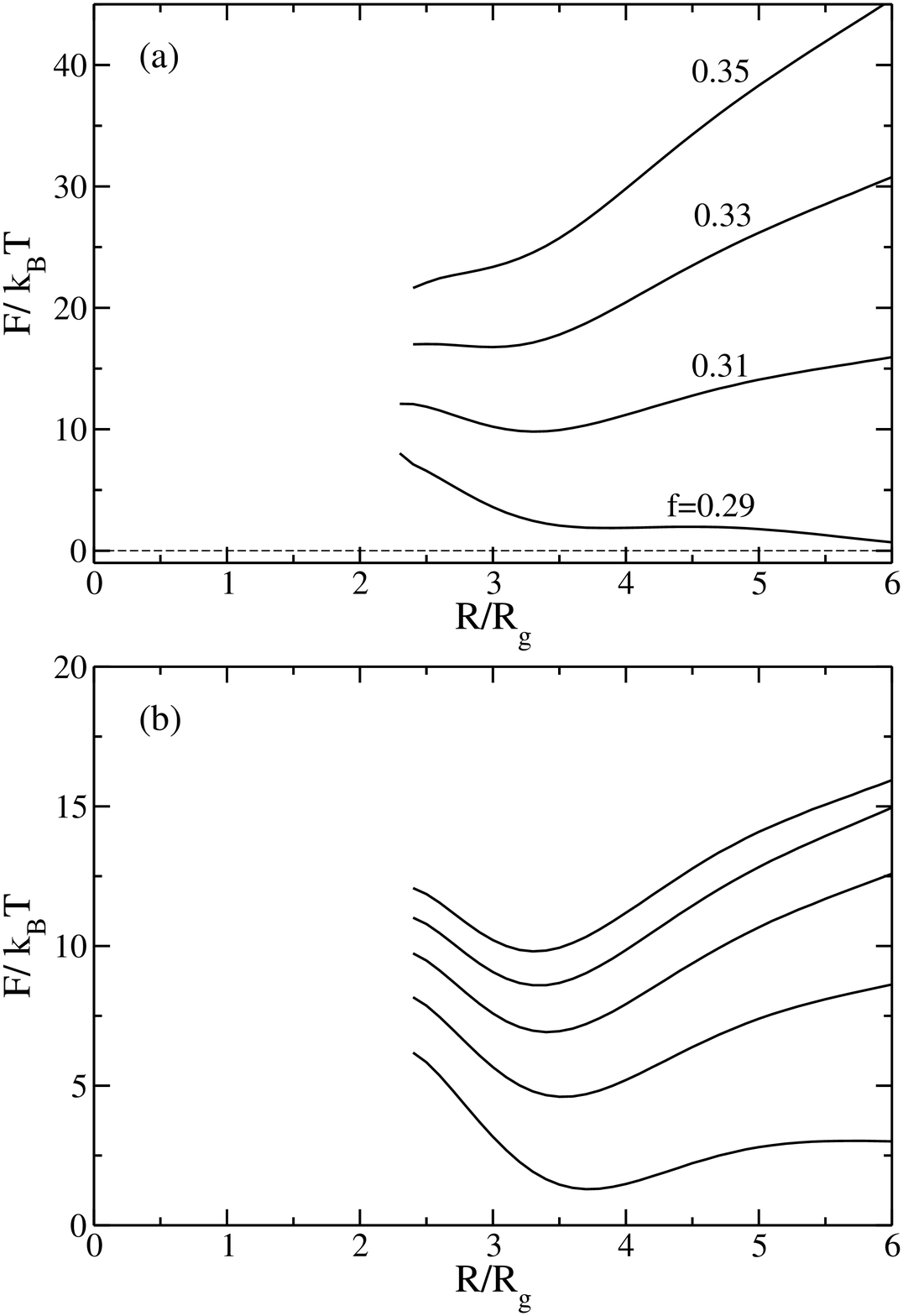}
\caption{(a) Free energy of an IMI 
as a function of $R/R_g$ at zero tension 
for various amphiphile architectures, $f$. (b) Free energy of an IMI 
with
$f=0.31$ and for various tensions $\gamma/\gamma_0$. From top to bottom,
$\gamma/\gamma_0$ varies from 0.0 to 0.4 in increments of 0.1. Minima on these
curves correspond to metastable IMI structures.}
\label{fig:imifree}%
\end{figure}
%%%%%%%%%%%%%%%%%%%%%%%%%%%%%%%%%%%%%

The free energy of the IMI can be calculated readily because it possesses
axial symmety. In Fig.~\ref{fig:imifree}(a) we show its free 
energy as a function of radius for a bilayer under zero tension for various
architectural parameters $f$. Again, as is 
the case with the other axially symmetric structures we studied, the 
free energy is asymptotically linear at large $R$, with the slope
$2\pi\lambda_{IMI}$ defining the effective line tension $\lambda_{IMI}$ 
({\em c.f.} Eq.~\ref{holefree}).
In Fig.~\ref{fig:imifree}(b) we also show the free energy of 
the IMI as a function of $R$ for a bilayer with fixed $f=0.31$ and different 
tensions. 
From \ref{fig:imifree} it is apparent that the free energy of the structure 
for the sizes that are pertinent to the fusion intermediate cannot be described 
by a simple estimate based on the line tension of the IMI. The increase of the
free energy with decreasing radius at small radius results from the
repulsion of the interfaces across
the IMI structure. It is similar to the free energy barrier associated
with closing the fusion pore \cite{Katsov04}.
Note that the free energy does not decrease with $R$ for large 
$R$ because the IMI does {\em not} eliminate  bilayer area. Therefore 
for large enough $\alpha$ and/or radius $R$, the free energy of this 
structure will exceed that of the stalk-hole complex in which a hole forms 
next to the elongated stalk. 
We turn now to the calculation of the free 
energy of this complex.

\subsection{Immediately after formation of stalk-hole complex}
 
%%%%%%%%%%%%%%%%%%%%%%%%%%%%%%%%%%%%%%%%
%%%%%%%%%%%%%%%%%%%%%%%%%%%%%%%%%%%%%%%%
%%%%%%%%% FIG 6
%%%%%%%%%%%%%%%%%%%%%%%%%%%%%%%%%%%%%%%%
%%%%%%%%%%%%%%%%%%%%%%%%%%%%%%%%%%%%%%%%

\begin{figure}[t]
\includegraphics[width=3in,angle=0]{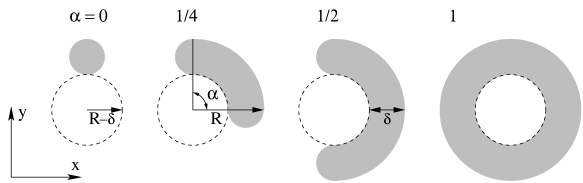}
\caption{Parametrization of the stalk-hole complex. Gray schematically shows
  location of the hydrophobic segments in the plane of symmetry between fusing
  bilayers. The arc radius $R$ corresponds to
  the radial distance to the hydrophilic/hydrophobic interface of the
  hemifusion intermediate in the plane of symmetry. Projection of the edge of
  a hole in one of the membranes is shown with dashed line. The radius of this
  hole is $R-\delta$. The other membrane does not
  have a hole. The hydrophobic thickness of the bilayer is $\delta$.
  Values of the fractional arc angle $\alpha$, defined in the range $[0,1]$, are
  given at the top of each stalk configuration. }
\label{fig:estalknhole}%
\end{figure}
%%%%%%%%%%%%%%%%%%%%%%%%%%%%%%%%%%%%%%%%

We model the stalk-hole complex as an elongated stalk
in contact with a circular hole in one of the bilayers. We assume that the
radial axes of the elongated stalk and of the hole coincide and that the
radius of the hole is $R-\delta$, where $R$ is the radius of the elongated
stalk, and that $\delta$  is chosen such that the hole is aligned in the radial
direction with the elongated stalk over a fraction of its circumference, again
denoted by $\alpha$. See Fig.~\ref{fig:estalknhole}.  
To calculate the free energy of this configuration, we note that 
at $\alpha=1$, the configuration is simply a hemifusion intermediate (HI) 
of radius $R$, and the elongated stalk would now connect two bilayers to 
one bilayer.  We have calculated the free energy of the hemifusion 
intermediate previously \cite{Katsov04}. The 
radius $R$ of this structure is defined by the position of the
hydrophilic/hydrophobic interface in the $z=0$ symmetry plane. With 
these definitions of the radii of the hemifusion intermediate and of the 
hole, a choice of $\delta(\gamma,f)$ equal to the hydrophobic thickness of 
a bilayer, ensures that the hole is adjacent to the elongated stalk. In 
general, when the hole forms, the elongated stalk does not completely 
surround it, so that a fraction $1-\alpha$ of the stalk-hole complex looks 
like a bare hole edge in an isolated bilayer. Thus we approximate the 
free energy of this stalk-hole complex to be
 \begin{equation}
 F_2(R,\alpha)=\alpha F_{HI}(R)+(1-\alpha)F_H(R-\delta)+F_d.
 \label{justafter}
 \end{equation} 
The free energy $F_d$ comes from the end caps of the 
elongated stalk connecting to the hole edge. 
The two ends together do not make an axially symmetric 
stalk, but like the stalk, this defect is also saddle-shaped so one 
expects its free energy to be small and not very different from that of 
the stalk. 

\subsection{The transition state}

It is clear from Eqs.~\ref{justbefore} and \ref{justafter} that the 
free energies of these structures depend both on the radius, $R$, of the 
intermediate and on the fraction, $\alpha$. Thus we must consider a 
two-dimensional reaction coordinate space, $(R,\alpha)$. 
The fusion process starts off by the formation of the classical stalk, which
corresponds to the $\alpha=0$ line on the $F_1(R,\alpha)$ free energy surface.
Elongation of the stalk corresponds to non-zero values of $\alpha$. 
We assume that the stalk-hole complex forms when the free energy surfaces
$F_1(R,\alpha)$ and $F_2(R,\alpha)$ intersect. 
This intersection happens along a line in the
$(R,\alpha)$ plane, which defines the ``ridge'' of possible transition states
$(R,\alpha_{TS}(R))$ with 
 \begin{eqnarray}
 \alpha_{TS}(R)&=&\frac{F_H(R-\delta)+
F_d-F_S}{F_H(R_\delta)+F_{IMI}(R)-F_{HI}(R)},\\
                &=&1-
      \frac{F_{IMI}(R)-F_{HI}(R)+F_S-F_d}{F_H(R-\delta)+
F_{IMI}(R)-F_{HI}(R)}
 \label{alphacomplex}
 \end{eqnarray} 
The free energy of the {\em optimal} transition state 
can obtained by finding the free energy minimum along the ridge of the
transition states, which we shall do momentarily. First we note from
Eq.~\ref{alphacomplex} that the fraction of the hole surrounded by the  
elongated stalk increases as the free energy of an isolated hole 
increases. This shows that the reduction of the 
high cost of the bare hole edge is a driving force of this mechanism.
 
%%%%%%%%%%%%%%%%%%%%%%%%%%%%%%%%%%%%%%%%%%%%%%%%%
%%%%%%%%%%%%%%%%%%%%%%%%%%%%%%%%%%%%%%%%%%%%%%%%%
%%%%%%%% Fig. 7
%%%%%%%%%%%%%%%%%%%%%%%%%%%%%%%%%%%%%%%%%%%%%%%%%
%%%%%%%%%%%%%%%%%%%%%%%%%%%%%%%%%%%%%%%%%%%%%%%%%

\begin{figure}[t]
\includegraphics[width=3in]{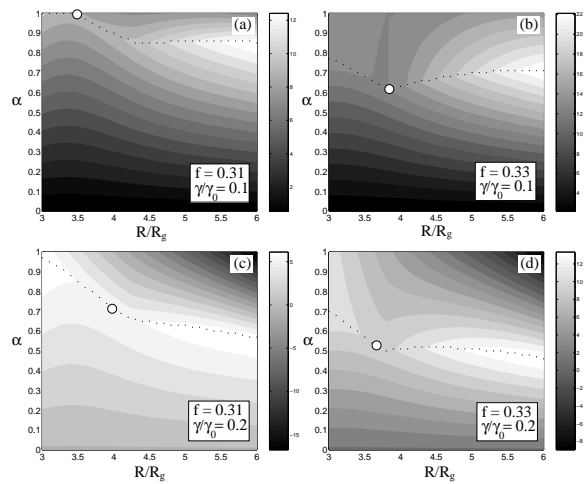}
  \caption{ Four free energy landscapes (in units of $k_BT$) of the fusion
    process, plotted as a function of the radius, $R$ (in units of
    $R_g$) and circumference fraction $\alpha$. The architecture of the
    amphiphiles and the value of the tension $\gamma/\gamma_0$ are given.
    The dotted line shows a ridge of possible transition states, separating
    two valleys. The region close to the $\alpha=0$ line corresponds to  
    a barely elongated stalk intermediate (see Eq.~\ref{justbefore}). 
    The other valley, close to $\alpha=1$ states, corresponds to a hole
    almost completely surrounded by an elongated stalk. The saddle point on the ridge,
    denoted by white dot, corresponds to the optimal (lowest free energy)
    transition state. The energy of the defect, $F_d$ has been set to zero
    here.}  
  \label{fig:landscape}%
\end{figure}
%%%%%%%%%%%%%%%%%%%%%%%%%%%%%%%%%%%%%%%%%%%%%%%%%

We return to the free energy landscape of the fusion process defined by 
${\rm min}(F_1(R,\alpha),F_2(R,\alpha))$. Examples of such landscapes are
shown in Fig.~\ref{fig:landscape}.
To clarify the effect of different parameters we present results for a membrane
consisting of lipids with $f=0.31$ and $0.33$, and under the reduced membrane
tension  $\gamma/\gamma_0=0.1$ and $0.2$. 
To obtain these results we have set the small defect energy $F_d$ to zero. 
This parameter has very little effect on the qualitative features of the
landscapes. Quantitative effects are also small and will be discussed below.

The landscapes are saddle-shaped with low free energy valleys close to $\alpha=0$
and $\alpha=1$ lines. The first valley corresponds to barely elongated stalks
of very small circumference, configurations which are clearly energetically 
inexpensive. The second valley
corresponds to a hole that is almost completely surrounded by the 
elongated stalk. Its energy is small because formation of the hole 
leads to a decrease of the membrane area under tension. One should note that
$\alpha=1$  corresponds to the hemifusion intermediate, which is also formed
in the standard mechanism, but through a completely different pathway.

The ridge of the transition states $(R,\alpha_{TS}(R))$  
is indicated by a dotted line. There is a saddle point along this ridge,
denoted by a circle on the plots.  
We denote the value of the radius of this optimal transition state as $R^*$, 
and the value of $\alpha_{TS}(R^*)$ as $\alpha^*$. The free energy 
of the transition state $F^*\equiv F_1(R^*,\alpha^*)=F_2(R^*,\alpha^*)$
This assumes that one can ignore any  
additional barriers caused by the rearrangement of amphiphiles in passing 
from the configuration just before the formation of the stalk-hole complex 
to the configuration just after. 
 
%%%%%%%%%%%%%%%%%%%%%%%%%%%%%%%%%%%%%%%%%%%%%%%%%%%
%%%%%%%%%%%%%%%%%%%%%%%%%%%%%%%%%%%%%%%%%%%%%%%%%%
%%%%%%%%%%%%%  Figure 8
%%%%%%%%%%%%%%%%%%%%%%%%%%%%%%%%%%%%%%%%%%%%%%%%%%
%%%%%%%%%%%%%%%%%%%%%%%%%%%%%%%%%%%%%%%%%%%%%%%%%%

\begin{figure}[t]
\includegraphics[width=3in]{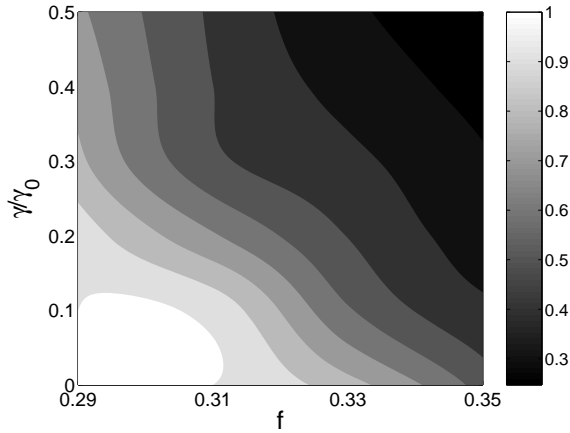}
\caption{ Plot of $\alpha^*$, which corresponds to the optimal transition
  state in the stalk-hole mechanism, as a function of architecture of the
  amphiphiles and the tension of the membrane.} 
\label{fig:alpha}%
\end{figure}
%%%%%%%%%%%%%%%%%%%%%%%%%%%%%%%%%%%%%%%%%%%%%%%%%%%

The value of $\alpha$ at the saddle point, $\alpha^*$, is shown in
Fig.~\ref{fig:alpha}. Once the stalk-hole
complex has formed, the free energy of the complex decreases as the stalk 
continues to enclose the hole, that is, as $\alpha$ increases to unity. 
This is clear {\em a priori} because as the stalk advances around the 
perimeter of the hole, it reduces the large line tension of the bare hole 
to the smaller line tension of the hole surrounded by stalk without any 
concommitent increase in energy due to that advance.

%%%%%%%%%%%%%%%%%%%%%%%%%%%%%%%%%%%%%%%%%%%%%%%%%%%%%%
%%%%%%%%%%%%%%%%%%%%%%%%%%%%%%%%%%%%%%%%%%%%%%%%%%%%%%
%%%%%%%%%%   Fig. 9
%%%%%%%%%%%%%%%%%%%%%%%%%%%%%%%%%%%%%%%%%%%%%%%%%%%%%%
%%%%%%%%%%%%%%%%%%%%%%%%%%%%%%%%%%%%%%%%%%%%%%%%%%%%%%

\begin{figure}[t]
\includegraphics[width=3in]{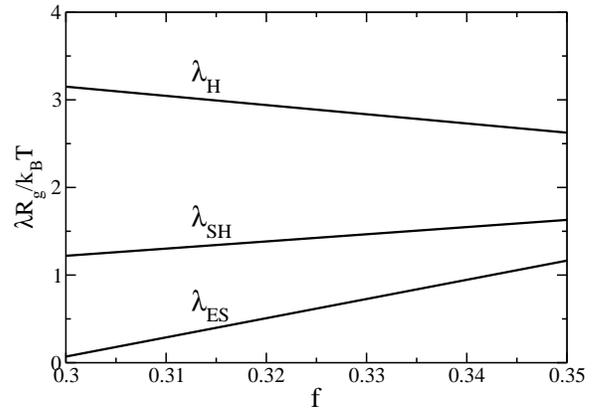}
\caption{Line tensions of an elongated linear stalk, $\lambda_{ES}$, of a bare 
hole in a membrane, $\lambda_H$, and of a hole which forms next to an
elongated stalk, $\lambda_{SH}$ as a function of architecture, $f$. 
All line tensions are in units of $k_BT/R_g$.
} 
\label{fig:tensions}%
\end{figure}
%%%%%%%%%%%%%%%%%%%%%%%%%%%%%%%%%%%%%%%%%%%%%%%%%%%%%%

For small values of the architectural parameter $f$, there 
is a considerable region for which $\alpha^*=1$. The reason for this can 
be inferred from Fig.~\ref{fig:tensions} which shows the calculated asymptotic
(large $R$) values of the line tension, $\lambda_{ES}$, of the elongated 
stalk. One sees that $\lambda_{ES}$ decreases as a function of $f$ so that
the free energy of an IMI,  
$F_{IMI}(R)$, which is dominated by this line tension, also decreases. 
Thus when the hole appears next to the extended stalk
in a membrane characterized by a small $f$, more of the hole 
will be surrounded by the stalk, that is, $\alpha$ will increase towards 
unity. This physical explanation is reflected in Eq.~\ref{alphacomplex}.

We expect that this result has consequences for the amount of transient 
leakage during the fusion event. It is reasonable to expect that the 
amount of leakage would decrease as $(1-\alpha)$, because this 
is the fraction of the hole in the stalk-hole complex which is not 
``sealed'' by the stalk. In fact, for architectures with sufficiently 
small $f$, ({\em i.e.} sufficiently large, negative spontaneous 
curvatures), Fig. \ref{fig:alpha} leads us to expect that $(1-\alpha)=0$ 
so that there would be no transient leakage at all. 

%%%%%%%%%%%%%%%%%%%%%%%%%%%%%%%%%%%%%%%%%%%%%%%%%%%%%%%
%%%%%%%%%%%%%%%%%%%%%%%%%%%%%%%%%%%%%%%%%%%%%%%%%%%%%%%
%%%%%%%%%   FIG 10
%%%%%%%%%%%%%%%%%%%%%%%%%%%%%%%%%%%%%%%%%%%%%%%%%%%%%%%
%%%%%%%%%%%%%%%%%%%%%%%%%%%%%%%%%%%%%%%%%%%%%%%%%%%%%%%

\begin{figure}[t]
\includegraphics[width=3in]{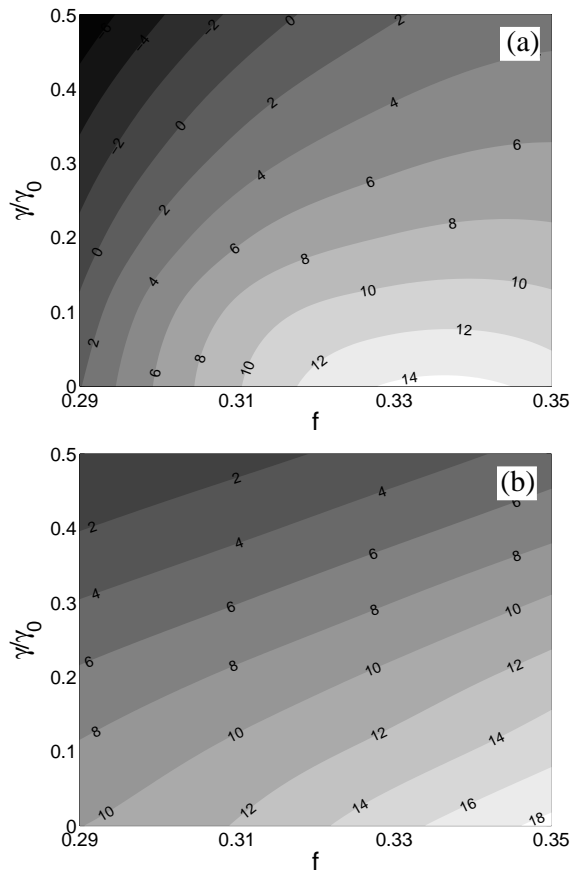}
\caption{Free energy barriers, in units of $k_BT$, in 
(a) the new stalk-hole complex mechanism, and
  (b) the standard hemifusion mechanism.}
\label{fig:FEcompare}%
\end{figure}
%%%%%%%%%%%%%%%%%%%%%%%%%%%%%%%%%%%%%%%%%%%%%%%%%%%%%%%

\begin{figure}[t]
\includegraphics[width=3in]{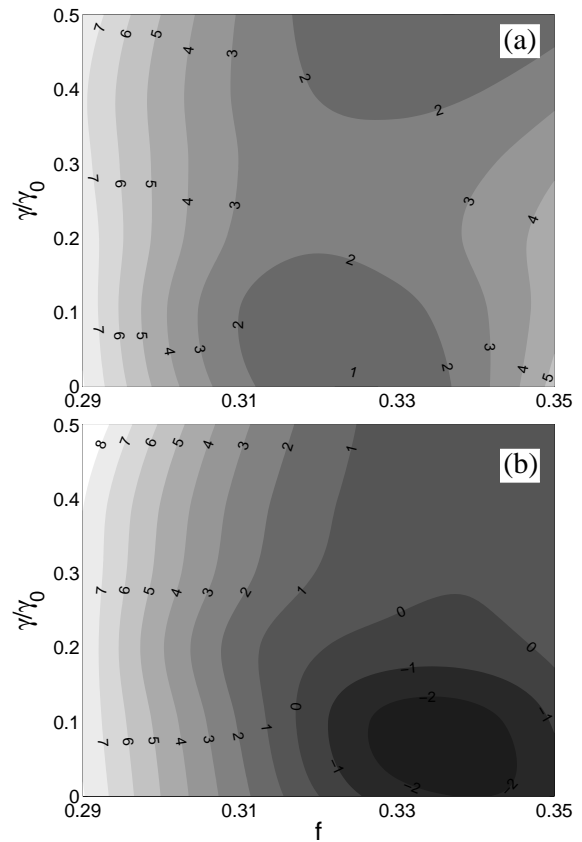}
\caption{(a) Difference between the free energy barrier in the standard
mechanism and that in the new mechanism, in units of $k_BT$, as a
function of architecture and tension. The defect free energy is here
taken to be zero. (b) Same as in (a) except that the defect free energy
is taken to be 4$k_BT$.}
\label{fig:difference}%
\end{figure}

The free energy barrier to formation of the stalk-hole complex  
measured relative to the initial metastable stalk, $(F^*-F_S)/k_BT$, is shown in 
Fig.~\ref{fig:FEcompare}(a). For comparison, we also show the barrier
encountered in the standard hemifusion expansion mechanism, which we
calculated earlier for the same parameters. 
It is clear that in both mechanisms the free energy
barrier can be significantly lowered either by an increase in the membrane
tension or decrease in the hydrophylic fraction $f$ (more negative spontaneous 
curvature).  The difference between these two barrier 
heights, in units of $k_BT$, is shown in Fig. \ref{fig:difference}. It is 
positive when the barrier in the old mechanism exceeds that of the new 
mechanism. We see that over the 
entire region, the barrier to fusion is lower in the new mechanism, and 
becomes increasingly favorable as $f$ decreases, {\em i.e.} as the 
amphiphile architecture becomes more inverted-hexagonal forming. We 
estimate that the difference in barrier heights in this system of block 
copolymers, from about 1 to 7$k_BT$, would translate to a range of 3 to 
18$k_BT$ in a system of biological lipids.
 
%%%%%%%%%%%%%%%%%%%%%%%%%%%%%%%%%%%%%%%%%%%%%%%%%%%%%%%
%%%%%%%%%%%%%%%%%%%%%%%%%%%%%%%%%%%%%%%%%%%%%%%%%%%%%%%
%%%%%%%%%   FIG 11
%%%%%%%%%%%%%%%%%%%%%%%%%%%%%%%%%%%%%%%%%%%%%%%%%%%%%%%
%%%%%%%%%%%%%%%%%%%%%%%%%%%%%%%%%%%%%%%%%%%%%%%%%%%%%%%

%%%%%%%%%%%%%%%%%%%%%%%%%%%%%%%%%%%%%%%%%%%%%%%%%%%%%%%

As noted earlier, we have set $F_d$ to zero in the above. Recall 
that the defect is the free energy of the two caps at the ends of the rim 
of an incomplete hemifusion diaphragm. As these caps are similar, but not 
identical, to the two halves of a stalk, we expect their energies to be 
similar. From our calculations \cite{Katsov04}, we know that the stalk 
free energy in our system does not exceed 4$k_BT$. If we set $F_d$ to 
4$k_BT$, then the difference in barrier heights in the two mechanism 
changes somewhat, and is shown in part (b) of Fig.~\ref{fig:difference}. 
The new mechanism is still favored over most of the tension/architecture 
space, while the standard mechanism is now favored for bilayers composed 
of amphiphiles with larger values of $f$ {\em i.e lamellar-forming lipids} 
and under small tensions.

\subsection{Formation of the final state}
\begin{figure}[t]
\includegraphics[width=3in]{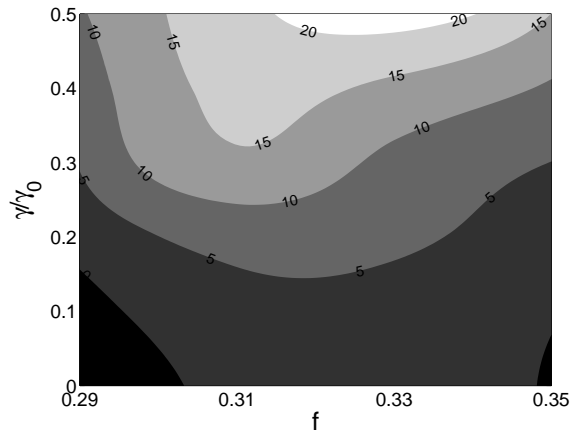}
\caption{Difference in free energy, in units of $k_BT$, between the stalk-hole transition state and
fusion pore of the same radius.}
\label{fig:FE_AM-pore}%
\end{figure}

The stalk-hole complex is a transition state along the fusion pathway,
but for complete fusion to occur it has to transforn into the fusion pore.
Properties of the fusion pore have been considered in detail in our first paper
\cite{Katsov04}. In the case of the hemifusion mechanism we have found that
the fusion pore has a lower free energy than the hemifused transition state
and thus can be formed without an appreciable additional barrier. In the
present case we also found that the fusion pore has a lower free energy than
the transition state everywhere except a very small region at small $f$ and
low tension, shown in black in Fig.~\ref{fig:FE_AM-pore}. This region of
parameters corresponds to the stalk-hole transition state with $\alpha=1$, 
{\em i.e.} a completely formed IMI (see Fig.~\ref{fig:alpha}). 
Therefore,
we conclude that formation of the stalk-hole complex involves the largest
free energy barrier along this pathway. 
In the special situation of very negative spontaneous
curvature amphiphiles and low tension,
the system does not continue on to the formation of a fusion pore,
but remains in a state in which the membranes are joined by an IMI
structure.
%This scenario is consistent with the fact that IMI structures have been
%experimentally observed.  

%%%%%%%%%%%%%%%%%%%%%%%%%%%%%%%%%%%%%%%%%%%%%%%%%%%%%%%
%%%%%%%%%%%%%%%%%%%%%%%%%%%%%%%%%%%%%%%%%%%%%%%%%%%%%%%
%%%%%%%%%   FIG 12
%%%%%%%%%%%%%%%%%%%%%%%%%%%%%%%%%%%%%%%%%%%%%%%%%%%%%%%
%%%%%%%%%%%%%%%%%%%%%%%%%%%%%%%%%%%%%%%%%%%%%%%%%%%%%%%

%%%%%%%%%%%%%%%%%%%%%%%%%%%%%%%%%%%%%%%%%%%%%%%%%%%%%%%

\section{Discussion}
\label{sec:disc}
We have utilized self consistent field theory and a model of polymeric 
bilayers to calculate the free energy barriers along the fusion pathway 
first seen by Noguchi and Takasu and by ourselves 
\cite{Noguchi01,Mueller02}.  There are at least two barriers associated 
with this path; a smaller one associated with the formation of the initial 
axially symmetric classical stalk, and a larger one associated with the formation of 
the stalk-hole complex. This path replaces the expensive step in the old 
mechanism, which is the radial expansion of the stalk into a hemifusion 
diaphragm, by the expensive step of elongating the stalk in a worm-like 
fashion and having a hole form next to it which creates the stalk-hole 
complex. There are several points that we wish to make.

First, by direct comparison of the calculated free energy barriers in the 
new mechanism and in the standard one, we have demonstrated that the free 
energy barriers are comparable. Hence this new pathway is a viable 
alternative to the standard mechanism. We have also demonstrated the 
tendency that the new mechanism tends to be the more favorable the more 
the amphiphile architecture approaches that of inverted-hexagonal formers.

Second, as noted previously, the new mechanism predicts the possibility of 
transient leakage {\em which is correlated in space and time} with fusion. 
Just such leakage, correlated in space and time with fusion, has been 
observed \cite{Frolov03}. This prediction is in contrast with the old 
mechanism in which any leakage that occurs is not correlated directly with 
the fusion process itself. Our calculations predict that the amount of 
this correlated leakage decreases, and can vanish altogether, as the 
architecture of the amphiphiles becomes more like that of 
inverted-hexagonal formers. This is a prediction that could be tested by 
carrying out a series of experiments like those of Frolov et al. 
\cite{Frolov03} on vesicles for which one could vary the amphiphile 
architecture or the relative composition of amphiphiles of different 
architecture. 
Such control of amphiphile architecture is readily obtained in 
polymersomes \cite{Discher99} which would
therefore offer an excellent system in which to test this prediction.

Third, our calculations predict existence of equilibrium IMI structures
that are metastable with respect to formation of the fusion pore, except in a
region of very low $f$ and $\gamma$ where they are actually favored over a pore. 
The possible occurrence of 
these structures had previously been dismissed due to
very high estimates of the free energy of their formation \cite{Siegel93}.

Finally we observe that in order for this new mechanism to be favorable, 
two conditions must be met. The first is that it must not cost too much 
free energy for the stalk to elongate in a worm-like fashion, in the manner 
that it does before the hole appears. That this can be the case is clear 
from the fact that at the transition to an inverted hexagonal phase, the 
line tension of linear stalks is small. Thus as the architecture is varied 
such that the system approaches this transition, it must be inexpensive 
for the stalk to elongate and wander. That this is correct can be seen 
from the calculated line tension, $\lambda_{ES}$, of the elongated linear stalk 
shown in Fig. \ref{fig:tensions}. It is essentially independent of 
tension, $\gamma$. We see that this line tension decreases with decreasing $f$ as 
expected, which decreases the cost of elongating a stalk. The second 
condition is that the free energy of the hole which is created must not be 
too large. As noted earlier, the high cost of an isolated hole is due to 
the line tension of its periphery. If this is reduced by causing the hole 
to form next to the elongated stalk, the cost of the hole in the 
stalk-hole complex will also be reduced. To determine whether this is so, 
we have calculated the line tension of an isolated hole in a bilayer, 
$\lambda_H$, and also the line tension of a hole created next to an 
elongated stalk, $\lambda_{SH}$. These results, again essentially 
independent of the membrane tension, are shown in Fig. \ref{fig:tensions}, 
as a function of architecture. It is seen that in the region of $f$ in 
which successful fusion is possible, $0.29<f<0.37$ \cite{Katsov04}, the 
line tension of the hole is reduced by about a factor of two. Let us now 
show that even such a relatively small change can have a very large effect 
on the rate of fusion.

Consider the simple estimate of the free energy of a hole,
Eq. \ref{holefree}, which we reproduce here
\begin{equation}
F_H=2\pi\lambda_H R-\pi\gamma R^2.
\end{equation}
The height of the barrier to stable hole formation corresponds to the 
maximum of this function. We ignore any $R$-dependence of $\lambda_H$ and 
$\gamma$ and immediately obtain the radius of the hole corresponding to 
the barrier to be $R^{*}=\lambda_H/\gamma$, and the height of the 
barrier to be $F^{*}=\pi\lambda_H^2/\gamma$. The rate of formation of an 
isolated hole in a bilayer is proportional to the Boltzmann factor
\begin{eqnarray}
P_H&=&\exp\{-[F^{*}-k_BT\ln (A_H/\ell^2)]/k_BT\},\\
        &=&\frac{A_H}{\ell^2}\exp(-\pi\lambda_H^2/\gamma k_BT),
\label{boltzmann}
\end{eqnarray}
where the entropy associated with the formation of a hole in an available 
area $A_H$ is $-k_B\ln(A_H/\ell^2)$ with $\ell$ a characteristic length on 
the order of the bilayer width. If $P_H\ll 1$, then the bilayer is stable to 
hole formation by thermal excitation.

The formation of the
stalk-hole complex reduces the line tension of that part of the 
hole near the stalk from $\lambda_H$ to $\lambda_{SH}$. 
This can be described by introducing the effective average line tension 
entering Eq.~\ref{boltzmann}  
\begin{equation}
\lambda_H\rightarrow
\bar\lambda_\alpha\equiv\alpha\lambda_{SH}+(1-\alpha)\lambda_H.
\end{equation}
Then the corresponding rate of stalk-hole complex formation becomes
\begin{equation}
P_{SH}=\frac{N_S a_S}{\ell^2}\exp\left(-\pi{\bar\lambda_\alpha}^2/\gamma k_BT
\right),
\end{equation}
where $N_S$ is the number of stalks formed in the system and $a_S$ is the area
around each stalk in which hole nucleation can take place.
For the small reduction $\lambda_{SH}/\lambda_H=1/2$, the above becomes
\begin{eqnarray}
\frac{P_{SH}}{P_H}&=&\frac{N_S a_S}{A_H}
\exp\left\{\frac{\pi\lambda_H^2}
{\gamma
k_BT}\left[\alpha\left(1-\frac{\alpha}{4}\right)\right]\right\},\\
&=&\frac{N_S a_S}{A_H}
\left(\frac{A_H}{\ell^2 P_H}\right)^{\alpha(1-\alpha/4)}.
\label{r1}
\end{eqnarray}
This shows explicitly that {\em if} the isolated membrane is stable to
hole formation, ({\em i.e.} $P_H\ll1$),  {\em then} even a small
reduction in the line tension
ensures that formation of the stalk/hole complex causes the rate of hole
formation in the apposed bilayers, and therefore fusion, to increase greatly.

We illustrate this with two examples. We first consider the copolymer 
membranes which we simulated previously \cite{Mueller02,Mueller03}. In 
that case the exponent in the Boltzmann factor
\begin{equation}
-\frac{\pi\lambda_{H}^2}{\gamma
k_BT}=-\pi\left(\frac{\lambda_H R_g}{k_BT}\right)^2\left(\frac{\gamma_0}
{\gamma}\right)\left(\frac{k_BT}{\gamma_0 R_g^2}\right),
\end{equation}
where $\gamma_0$ is the tension of an interface between bulk hydrophilic and
hydrophobic hompolymer phases. The various
factors in the simulated system are $\lambda_H R_g/k_BT=2.6$ at $f=0.35$ (see
Fig. \ref{fig:tensions}),  and $\gamma_0/\gamma=4/3$,
$k_BT/\gamma_0 R_g^2=0.31$, $A_H/{\ell^2}=39$
\cite{Mueller02,Mueller03}. From this we obtain
$P_H\approx 6\times 10^{-3}$, so that
isolated bilayers should have been stable to hole formation, as was
indeed the case. However in the presence of a stalk, the Boltzmann
factor will be increased according to Eq. \ref{r1}. If we assume that
the elongated stalk enclosed one half of the perimeter of the hole when
it appeared, ({\em i.e.} $\alpha=1/2$), and that $n_S a_S/A_H\sim 0.3$
% MM
(consistent with the simultaneous observation of multiple stalks in a small
simulation cell \cite{Mueller02}),  
we find that $P_{SH}/P_H\sim 14$
so that the rate of hole formation should have increased
appreciably as observed in the simulations.

This increase is expected to be more dramatic in biological membranes. In 
that case we estimate the exponent of the Boltzmann factor, 
$-\pi\lambda_H^2/\gamma k_BT$, as follows. We take the line tension to be 
that measured in a stearoyloleoylphosphatidylcholine and cholesterol 
bilayer, $\lambda_H\approx 2.6\times 10^{-6}$ erg/cm 
\cite{Zhelev93,Moroz97}. For the surface tension, we take an estimate of 
the energy released by the conformational change of four of perhaps six 
hemagglutinin trimers aranged around an area of radius 4nm, each trimer 
giving out about 60$k_BT$ \cite{Kozlov98}. This yields an energy per unit 
area $\gamma\approx$ 20 erg/cm$^2$. Thus $P_H=1.7\times 
10^{-11}(A_H/{\ell^2})$, which indicates that even subject to this large, local,
energy per unit area, the membrane is quite stable to hole formation for vesicles 
of any reasonable size. However if we assume again that the line tension 
of the hole is reduced by a factor of two by being nucleated next to the 
elongated stalk, that the stalk extends halfway around the circumference 
of the hole, and the density of stalks is such that $n_S a_S/A_H=0.3$, then 
the rate of hole formation is increased by 
\begin{eqnarray} 
\frac{P_{SH}}{P_H}&=&0.3\left(\frac{1}{1.7\times 10^{-11}}\right)^{7/16}, 
\nonumber\\
               &\sim& 1\times 10^4,
\end{eqnarray}
or an increase of more than four orders of magnitude. 

One should note the implications of this simple argument. Because the 
probability to form a stable hole depends exponentially on the square of 
the line tension, an isolated bilayer is guaranteed to be stable against 
hole formation for normal line tensions. However {\em it is precisely this 
same dependence which also ensures that the bilayer will be destabilized 
by hole formation due to any mechanism which even modestly reduces that 
line tension.} From here it is only a short step to successful fusion.

We acknowledge very useful conversations and correspondence 
with M. Kozlov and S.J. Marrink. This work was supported by the National
Science Foundation under Grant No. 0140500. Additional support was
provided by the Volkswagen foundation.

\bibliography{kkfusion04}

\end{document}